\documentclass[11pt,a4paper]{article}

\usepackage[a4paper,margin=2.7cm]{geometry}
\usepackage[T1]{fontenc}
\usepackage[utf8]{inputenc}

\usepackage{mathpazo} 

\usepackage{setspace}
\usepackage{microtype}
\usepackage{titlesec}
\usepackage{fancyhdr}
\usepackage{graphicx}
\usepackage{xcolor}
\usepackage{epigraph}
\usepackage{enumitem}
\usepackage{amsmath,amssymb}
\usepackage{hyperref}
\usepackage{csquotes}
\usepackage{caption}
\usepackage{booktabs}
\usepackage{abstract}
\usepackage{tabularx}

\definecolor{inonublue}{RGB}{28,52,91}
\definecolor{softgray}{RGB}{90,90,90}

\hypersetup{
    colorlinks=true,
    linkcolor=inonublue,
    urlcolor=inonublue,
    citecolor=inonublue
}

\titleformat{\section}
{\Large\bfseries}
{\thesection.}{0.5em}{}

\titleformat{\subsection}
{\large\bfseries}
{\thesubsection.}{0.5em}{}

\title{
    \vspace{-1.5cm}
    {\Huge\bfseries Erdal İnönü at 100:\\[0.1cm] From the Sphere to the Plane} \\
    \vspace{0.4cm}
    {\Large An Introduction to the İnönü-Wigner Contraction\\ and the Legacy of Erdal İnönü\footnote{This paper is based on a talk prepared for the İnönü-Barut-100 Workshop, to be delivered on June 7, 2026, at Boğaziçi University, Istanbul, Türkiye.}}
}

\author{
    \large \textbf{Ilmar Gahramanov} \\ [0.1cm]
    \small Department of Physics, Bogazici University, \\
    \small  34342 Bebek, Istanbul, Türkiye
}

\date{\today}

\begin{document}

\maketitle

\vspace{-0.7cm}

\begin{abstract}
On the centennial of Erdal İnönü's birth, this article reflects on his scientific legacy and his role in shaping modern theoretical physics in Türkiye. We briefly review his life, scientific vision, and contributions to the development of research institutions and the scientific community. Particular attention is devoted to his most celebrated scientific contribution, the İnönü-Wigner contraction, which established deep connections between different symmetry groups and became an important tool in modern theoretical physics. Using a simple geometric example, we present an accessible introduction to this idea and illustrate its significance for our understanding of physical theories.
\end{abstract}

\vspace{0.5cm}


\begin{center}
\includegraphics[width=0.70\textwidth]{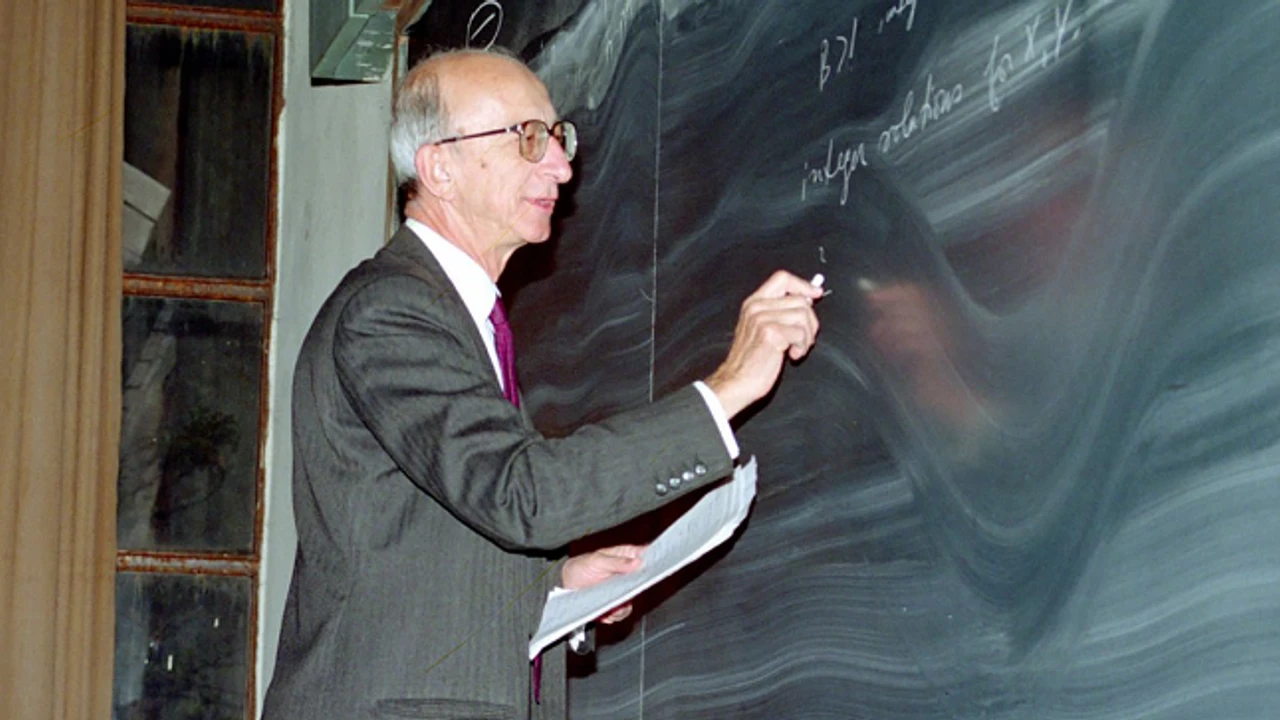}

\vspace{0.3cm}
\small\textit{Erdal İnönü lecturing at the blackboard. Throughout his career, he remained deeply committed to teaching, research, and the cultivation of scientific culture in Türkiye.}
\end{center}

\section*{Erdal İnönü and the Spirit of Theoretical Physics}

\epigraph{
Ne içindeyim zamanın,\\
Ne de büsbütün dışında;\\
Yekpâre, geniş bir ânın\\
Parçalanmaz akışında.

\vspace{0.5em}

\emph{Neither am I inside time,\\
Nor altogether without;\\
In the unbroken flow of\\
An instant singular and vast.
}
}
{\textit{Ahmet Hamdi Tanpınar}\\
\textit{(translation by Erdağ Göknar)}}

One hundred years after his birth, Erdal İnönü remains one of the central figures in the history of Turkish science. To summarize Erdal İnönü’s life very briefly, he was born on June 6, 1926, in Ankara. He studied physics at Ankara University and later pursued graduate studies at Caltech. After a brief stay at the Institute for Advanced Study in Princeton, he returned to Türkiye, where he worked at Ankara University and later at Middle East Technical University, serving first as Dean of the Faculty of Arts and Sciences and subsequently as Rector. In 1974, Erdal İnönü left Middle East Technical University and joined Boğaziçi University, where he held several administrative positions, including Chair of the Physics Department and Dean of the Faculty of Basic Sciences. In addition, he served as the founding director of the Basic Sciences Research Institute of the Scientific and Technological Research Council of Türkiye (TÜBİTAK) in Gebze\footnote{The institute was later relocated to Kandilli and renamed the Feza Gürsey Institute. Following its closure in 2011, its research activities were transferred back to Gebze.} and as President of the Turkish Physical Society. In the later part of his life, he left academia and entered politics. In the mid-1990s, following the end of his political career, Erdal İnönü returned to academic life, continued his scholarly work at the Feza Gürsey Institute, and taught courses on the history of science at Sabancı University.  Among the many distinctions he received, İnönü was awarded the TÜBİTAK Science Prize in 1974 and the Wigner Medal in 2004.

Inönü's talent revealed itself in many ways, above all in his remarkable ability to recognize, within ideas and developments only beginning to emerge in science and in the life of the country, what could become most valuable and important for the future. His gift lay not only in such foresight, but also in his determination to transform promising ideas into reality. Erdal İnönü was one of the principal initiators behind the TÜBİTAK Basic Sciences Research Institute, which later evolved into the Feza Gürsey Institute. The institute was established according to principles that were highly innovative for its time and played a crucial role in the development of fundamental research in Türkiye. İnönü also played a key role in reactivating the Turkish Physical Society and strengthening the country's scientific community.

At a time when theoretical physics in Türkiye was still developing, İnönü's presence gave younger researchers confidence that internationally recognized science could be created within the country itself. Although Erdal İnönü had few direct students\footnote{He supervised several students working on neutron transport theory \cite{inonu1970orthogonality}. Unfortunately, I have been unable to identify the names of these students or locate the titles of their theses and publications.} in the conventional sense of the term, his influence on educating and inspiring new generations of physicists was immense, extending far beyond his immediate academic circle \cite{Inonu1973TurkFizigi,Inonu1976SayisalGozlemler,Inonu1976IstanbuldaFizik}. The unique scientific atmosphere he fostered within the institutions where he worked, including Middle East Technical University, Boğaziçi University, and Feza Gürsey Institute, became an important source of intellectual and scientific formation for many young researchers.

His way of thinking was so distinctive that, for many of those around him, his intellectual vision often seemed unconventional and was not always immediately understood or appreciated. Reflecting on Erdal İnönü’s early period in the Physics Department at Boğaziçi University, Mahmut Hortaçsu remarks \cite{Hortacsu2026ErdalBey} that İnönü introduced a culture of research into a department whose activities had been centered largely on teaching. As might be expected, this shift in academic priorities was not universally welcomed, and his efforts were met with a degree of resistance:
\begin{displayquote}
{\itshape
Whenever we met Erdal İnönü, he would first ask the members of the Physics Department and later, after becoming dean: ``What have you done in your research this week? How much progress have you made?''

In this way, he helped establish a culture in which good teaching alone was not considered sufficient; active engagement in research became an essential part of academic life, at least within the department.
}
\end{displayquote}

His scientific talent, combined with exceptional personal qualities and broad intellectual interests, earned him the respect and admiration of generations of students, colleagues, and collaborators. In memoirs and tributes written after his passing \cite{Hortacsu2026ErdalBey,Hacinliyan2009ErdalInonu}, he is remembered not only as an accomplished physicist and statesman, but also as a person of remarkable humility, integrity, generosity, and learning. İnönü possessed the rare ability to think and act on many levels at once, with a genuine sense of responsibility toward the country and its scientific future. 

Deeply devoted to science, he nevertheless remained remarkably modest, almost shy, attentive and gentle in his manner, while at the same time being persistent and determined whenever it came to matters important for the nation. This is how Erdal İnönü was known to so many people, and this is also how he appears to us, even to those who never had the chance to meet him personally\footnote{I would like to acknowledge the many people with whom I discussed Erdal İnönü, as I never had the opportunity to meet him personally. My first visit to the Feza Gürsey Institute was in the summer of 2008, one year after his passing. Much of what I know about Erdal İnönü has been learned through the recollections and insights of those who had the privilege of knowing him. I am particularly grateful to Mahmut Hortaçsu, Metin Arık, Yani Skarlatos, Hasan Gümral and Teoman Turgut for generously sharing their memories and perspectives.}.

\begin{center}
\includegraphics[width=1.0\textwidth]{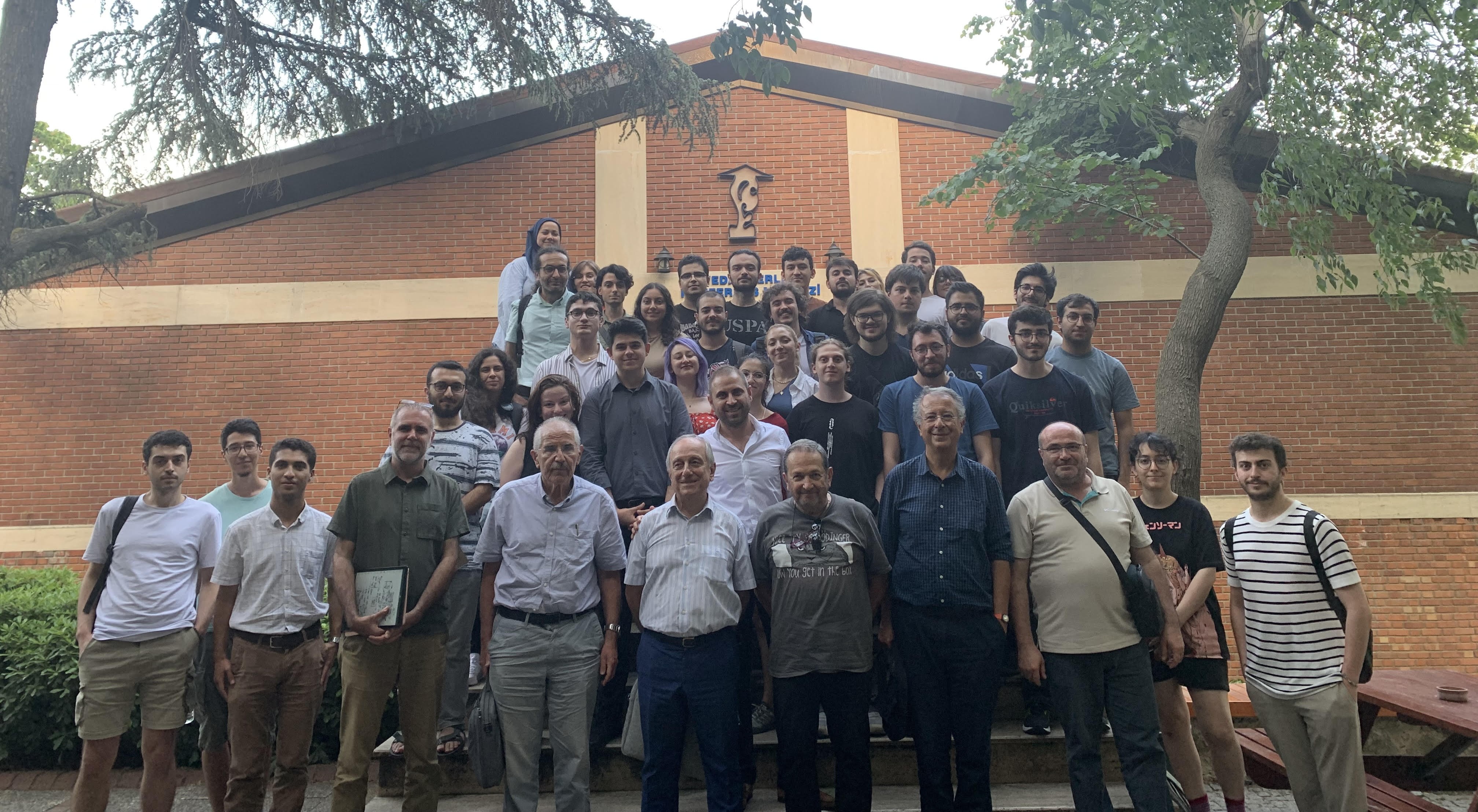}

\vspace{0.3cm}
\small\textit{Participants of the Workshop Dedicated to the Memory of Erdal İnönü,\\ Boğaziçi University, June 6, 2024.}
\end{center}

As the years pass, we are beginning to appreciate more fully, not only emotionally but also intellectually, the magnitude of the loss suffered by the Turkish community of fundamental scientists. For younger physicists, his legacy continues to serve both as inspiration and as a reminder of what an academic life can ideally represent.

Perhaps İnönü's greatest legacy lies not in any single scientific result, institution, or public office, but in the scientific culture he helped create and the generations of researchers shaped by it.

\section*{The Idea of an İnönü--Wigner Contraction}

\epigraph{
\hspace*{-1cm}\parbox{\dimexpr\textwidth+1cm\relax}{

I have serious reason to believe that the planet from which the little prince came is Asteroid B-612. This asteroid was seen only once through the telescope. That was by a Turkish astronomer, in 1909.
}
}{
\hspace*{-1cm}\textit{Antoine de Saint-Exupéry, \emph{The Little Prince}}
}

One of İnönü’s most influential contributions is the celebrated İnönü-Wigner contraction\footnote{One should also mention the work of Segal \cite{segal1951class}, which proposed a more general limiting process between non-isomorphic groups.} \cite{Inonu:1953sp} (see also \cite{inonu1956some,inonu2003fiftieth,inonu1997historical}), developed jointly with Eugene Wigner. This idea established connections between different symmetry groups and became an important tool in both mathematics and theoretical physics.In the İnönü-Wigner approach, a continuous parameter is introduced such that, in an appropriate limit, the structure constants of the Lie algebra are deformed while the dimension of the group remains unchanged. It provides a mathematical realization of the principle that, whenever a new theory generalizes an older one, there should exist a well-defined limiting procedure through which the results of the older theory are recovered.

A simple example can be understood using the geometry of a sphere \cite{Gilmore1974,Knap:2023cgk,Kim:1996wa,bacskal2024group}. Consider a three-dimensional space with axes \(x\), \(y\), and \(z\). There are three basic rotations: around the \(x\)-axis, around the \(y\)-axis, around the \(z\)-axis. These rotations generate all other rotations. Therefore, one introduces three generators $ J_x, J_y$ and $J_z$. The difference between performing two operations in different orders is described by the commutator
\[
[J_x,J_y]=J_xJ_y-J_yJ_x.
\]
For rotations one finds
\[
[J_x,J_y]=J_z,
\qquad
[J_y,J_z]=J_x,
\qquad
[J_z,J_x]=J_y.
\]
The meaning is that  if one performs an infinitesimal rotation about the \(x\)-axis and then about the \(y\)-axis, and compares the result with performing the same rotations in the opposite order, the difference is equivalent to an infinitesimal rotation about the \(z\)-axis. In other words, infinitesimal rotations about different axes do not commute, and their failure to commute produces a rotation about the third axis (see, e.g. \cite{Gursey:1964zza,raczka1986theory}).

Now imagine standing near the North Pole of the sphere. A small rotation about the $x$-axis moves us slightly in one horizontal direction, while a small rotation about the $y$-axis moves us in the perpendicular direction. To describe these small motions, define
\[
P_x=\frac{J_x}{R},
\qquad
P_y=\frac{J_y}{R}.
\]
These rescaled generators may be interpreted as infinitesimal displacements on the surface of the sphere. As the radius becomes larger and larger, the sphere appears flatter and flatter in a neighborhood of the North Pole. In the limit $ R\to\infty,$ the sphere becomes indistinguishable from a plane.

Let us now examine what happens to the algebra of rotations when the sphere becomes very large.  The idea is that, on a very large sphere, a small rotation around the $x$- or $y$-axis moves a point only slightly along the surface. Such motions begin to resemble translations on a flat plane. The factors of $1/R$ are introduced precisely to keep these motions finite in the large-radius limit.

We can make this idea precise by computing the commutator of the new generators
\[
[P_x,P_y]
=
\frac{1}{R^2}[J_x,J_y]
=
\frac{J_z}{R^2}.
\]
As the radius of the sphere grows, the factor $1/R^2$ becomes smaller and smaller. Consequently, $[P_x,P_y]\rightarrow 0$ when $R\to\infty$. In the limit of an infinitely large sphere we therefore obtain
\[
[P_x,P_y]=0.
\]
This result has a simple geometric interpretation. On a plane, translating first in the $x$-direction and then in the $y$-direction leads to exactly the same final position as performing the translations in the opposite order. The corresponding generators therefore commute. We see that the rescaled rotations $P_x$ and $P_y$ acquire precisely this property in the large-$R$ limit.

Next, let us examine the commutators involving $J_z$. Using the original rotation algebra, one finds
\[
[J_z,P_x]=P_y,
\qquad
[J_z,P_y]=-P_x.
\]
These relations tell us that the generators $P_x$ and $P_y$ transform into one another under rotations generated by $J_z$. This is exactly what happens to ordinary translations in a two-dimensional plane: a rotation changes the direction of a translation vector but does not destroy its translational character.

Collecting the limiting commutation relations,
\[
[P_x,P_y]=0,
\qquad
[J_z,P_x]=P_y,
\qquad
[J_z,P_y]=-P_x,
\]
we recognize the Lie algebra of the two-dimensional Euclidean group $E(2)$. This is the symmetry group of the ordinary plane, consisting of rotations and translations. Therefore we obtain
\[
SO(3)
\quad
\xrightarrow{\;R\to\infty\;}
\quad
E(2).
\]
This limiting procedure is called an \emph{İnönü-Wigner contraction}. Geometrically rotations that move a point slightly along the surface of a very large sphere become ordinary translations on a flat plane.

A historically important example arises in the transition from Einstein's special relativity to Newtonian mechanics. The symmetry group of special relativity is the Poincar\'e group \cite{Landau:1975pou}. The symmetry group of Newtonian mechanics is the Galilei group \cite{Landau:1972}. When the speed of light becomes infinitely large, $c\to\infty,$ relativistic effects disappear and one obtains
\[
\text{Poincar\'e Group}
\quad
\xrightarrow{\;c\to\infty\;}
\quad
\text{Galilei Group}.
\]
Many other examples are known today, and contractions have become a standard tool for understanding how different physical theories are related through limiting procedures.

The sphere becoming a plane provides a particularly simple illustration of the idea introduced by İnönü and Wigner more than seventy years ago. Despite its simplicity, it captures the essence of the contraction concept in an especially transparent way. Today, the İnönü–Wigner contraction is a standard topic in textbooks on Lie groups and Lie algebras and remains an important tool in many areas of theoretical physics \cite{saletan1961contraction,Wybourne1974,weimar2000contractions,fialowski2005deformations}. The enduring influence of this idea on modern physics reminds us that Erdal İnönü's legacy lives not only in the institutions he helped build and the generations he inspired, but also in the mathematical ideas that continue to shape our understanding of nature.

\section*{Acknowledgments}

I would like to thank all the organizers of the Inönü-Barut-100 Workshop, organized by the Istanbul Integrability and Stringy Topics Initiative (istringy.org), for their efforts in making this event possible. I am particularly grateful to İsmail Kılıç, Reyhan Yumuşak, Şahin Çetin and Doğa Sağlam for their dedication and assistance.


\nocite{*}
\bibliographystyle{unsrt}
\bibliography{references}

\end{document}